\documentclass[11pt,titlepage]{article}
\usepackage{amsmath}
\usepackage{amsfonts}
\usepackage{amssymb}
\usepackage{array}
\usepackage{graphicx}
\usepackage{float}
\usepackage{color}
\usepackage{ulem}
\usepackage[round,numbers,sort&compress]{natbib}
\usepackage{authblk}

\begin{document}
\title{Structural characterization and statistical-mechanical model of epidermal patterns}

\author[1]{D. Chen}
\author[2]{W. Y. Aw}
\author[2]{D. Devenport}
\author[1,3,4,5]{S. Torquato\footnote{torquato@electron.princeton.edu}}

\affil[1]{Department of Chemistry, Princeton University, Princeton NJ 08544}
\affil[2]{Department of Molecular Biology, Princeton University, Princeton NJ 08544}
\affil[3]{Department of Physics, Princeton University, Princeton NJ 08544}
\affil[4]{Princeton Institute for the Science and Technology of Materials, Princeton University, Princeton NJ 08544}
\affil[5]{Program in Applied and Computational Mathematics, Princeton University, Princeton NJ 08544}

%\author{D. C.}

%\affil{Department of Chemistry, Princeton University, Princeton NJ 08544}

%\author{W. Y. A.}

%\affiliation{Department of Molecular Biology, Princeton University, Princeton NJ 08544}

%\author{D. D.}

%\email{danelle@princeton.edu}

%\affil{Department of Molecular Biology, Princeton University, Princeton NJ 08544}

%\author{D. Chen, W. Y. Aw, D. Devenport, S. Torquato}

%\email{torquato@electron.princeton.edu}

%\affiliation{Department of Chemistry, Department of Physics, Princeton Institute for the Science and Technology of Materials, and Program in Applied and Computational Mathematics, Princeton University, Princeton NJ 08544}

% Revision date - uncomment to exclude date in the final version
\date{}

% Running head
\pagestyle{myheadings}
\markright{Statistical Mechanics of Skin Patterns}

% We are done with the headers, the actual document starts here

% generate the title page from the info in the headers above
\maketitle

% 200 words max Abstract
\begin{abstract}
In proliferating epithelia of mammalian skin, cells of irregular polygonal-like shapes pack into complex nearly flat two-dimensional structures that are pliable to deformations. In this work, we employ various sensitive correlation functions to quantitatively characterize structural features of evolving packings of epithelial cells across length scales in mouse skin. We find that the pair statistics in direct space (correlation function) and Fourier space (structure factor) of the cell centroids in the early stages of embryonic development show structural directional dependence (statistical anisotropy), which is a reflection of the fact that cells are stretched, which promotes uniaxial growth along the epithelial plane. In the late stages the patterns tend towards statistically isotropic states, as cells attain global polarization and epidermal growth shifts to produce the skin's outer stratified layers. We construct a minimalist four-component statistical-mechanical model involving effective isotropic pair interactions consisting of hard-core repulsion and extra short-ranged soft-core repulsion beyond the hard core, whose length scale is roughly the same as the hard core. The model parameters are optimized to match the sample pair statistics in both direct and Fourier spaces. By doing this, the parameters are biologically constrained. In contrast with many vertex-based models, our statistical-mechanical model does not explicitly incorporate information about the cell shapes and interfacial energy between cells; nonetheless, our model predicts essentially the same polygonal shape distribution and size disparity of cells found in experiments as measured by Voronoi statistics. Moreover, our simulated equilibrium liquid-like configurations are able to match other nontrivial unconstrained statistics, which is a testament to the power and novelty of the model. The array of structural descriptors that we deploy enable us to distinguish between normal, mechanically deformed, and pathological skin tissues. Our statistical-mechanical model enables one to generate tissue microstructure at will for further analysis. We also discuss ways in which our model might be extended so as to better understand morphogenesis (in particular the emergence of planar cell polarity), wound-healing, and disease progression processes in skin, and how it could be applied to the design of synthetic tissues.
\end{abstract}

% New page
\clearpage

\section*{Introduction}
Particle packing problems have great relevance not only in condensed-matter physics and mathematics \cite{Pa10, To10a, Co98, Co03}, but also in many biological contexts \cite{Mi92, El01, Ge08, Ji11, Ji14, Ta80, Ga16, Ch14b, Zh15}. Such examples include molecular ``crowding'' within cells \cite{Mi92, El01}, packings of living cells that comprise a variety of tissues \cite{Ge08, Ji11, Ji14}, and competitive settlement of territories by animals \cite{Ta80}, to name a few. In particular, many biological functions of animal tissues rely on the accurate formation of complex cell-packing patterns \cite{Hi08}. For example, it has been established \cite{Ji14} that in order for the avian retinae to sample light efficiently, photoreceptor cells pack into exotic disordered ``hyperuniform'' states \cite{To03}, in which (normalized) infinite-wavelength density fluctuations vanish. Disruption of cell packing patterns may lead to pathological conditions. For instance, it has been demonstrated \cite{Ji11} that packings of brain glioma cells possess the large-scale spatial correlations that are not observed in packings of benign brain white matter cells. In healthy cornea, cells pack into a disordered pattern that is transparent to visible light \cite{Ha69}, while cornea edema alters the cell-packing pattern, leading to blurry vision \cite{Ca84}.

During the last two decades, much work has been devoted to studying different aspects of epithelial patterns due to their importance for a variety of biological functions \cite{Gi06, De08, De11, De14, Ai10, Fa07, Hi08, St10, Ma07, Na13, Su09, Ad10, Ab12}. Epithelia are layers of cells that line the surfaces of organs and cover the exterior of the animal body. They serve as diffusion barriers that separate different physiological environments, protect the body from water loss, and prevent the permeation of toxins and pathogens \citep{Gi09, Si11, Gu13, Vr13}. They play a significant role in many biological processes such as embryonic development, organogenesis, homeostasis maintenance, and disease progression \cite{Gi09, Vr13}.

Epithelia can be classified into two categories by thickness: simple ones consisting of one single layer of cells, and stratified ones that have multiple layers \cite{Gi09, Vr13}. Within one layer, epithelia possess almost flat two-dimensional structure consisting of space-filling cells with nearly polygonal shapes \cite{Gi06}. Moreover, many epithelia display planar cell polarity (PCP), where cell polarity and structural features collectively align across the tissue plane, as exemplified by the uniform hair follicle alignment in the mammalian skin \cite{St09, Go11, De08, De11, De14, Ai10}. PCP is essential in various processes such as vertebrate gastrulation, mammalian ear patterning and hearing, and neural tube closure, to name a few \cite{De14, He00, Mu01, Co13}. In proliferating epidermis, cells pack themselves in a complex disordered fashion such that the epidermis are pliable to deformations \cite{Gi06}, which stands in contrast to those cells with regular hexagonal shapes that are arranged in an ordered fashion in insect retina and nonproliferating epithelia \cite{Re76, Lu11}.

Despite the extensive research works about epithelia, the underlying mechanisms for cells to assemble into complex tissue-level patterns in epithelia remain poorly understood \cite{Hi08}. For example, in cultured keratinocytes it is difficult to reconstruct certain \textit{in-vivo} structural features like PCP in skin \cite{De14}. Computational modeling \cite{Ho04, Fa07, Hi08, St10, Ma07, Su09, Ad10, Ab12, Jo15} provides a powerful means to shed light on this issue due to its relatively low cost and flexibility. In particular vertex-based models \cite{Fa07, Hi08, St10, Na01, Fl14} have been extensively employed to investigate epithelial patterns. Such models generally approximate cells as polygons or polyhedra and take into account cell elasticity and surface tension via certain energy functional forms \cite{Fa07, Hi08, St10}. These vertex models attempt to explain the formation of individual cell shapes, and other local properties, such as the coordination number (the number of near neighbors) and area distributions of cells. Recently there also have been models \cite{Li14, Bi16} that treat cell centroids as interacting particles with biologically-motivated cell mechanical terms to investigate cell motions in biological tissues.

In this work, our goal is to investigate both local and large-scale correlations of cell centroids in skin by exploiting the machinery of statistical mechanics \cite{Fr01, To02, To10a, Ha86, Ch00, Re05, Re06a, Re06b, Ma11, Ma13, Ja13, Zh13, Za09} to quantitatively characterize the structure of tissue samples and treat the cells as interacting entities with effective pair potentials \cite{Le00, Sz06, Be08, He11, Ba11, Se13, Ba13, Ga15, Ji14}. Specifically we first acquire histological images of mouse epidermal patterns in different stages of embryonic development. We then compute the pair correlation function and structure factor \cite{To02} of the point configurations associated with the cell centroids in these patterns. These pair statistics reveal a strong structural directional dependence (statistical anisotropy) across length scales at early stages, which is a reflection of the fact that cells are stretched to promote uniaxial growth \cite{In06, Aw16}. By contrast, the cell patterns in the late stages evolve to statistically isotropic states. This switch coincides with the establishment of global cell polarization and a shift to stratified growth in the developing epidermis \cite{De11, Le05, Aw16}. The pair statistics that we employ, capture the spatial correlations of cell centroids across length scales that are not reflected by single-cell statistics, such as cell size and density \cite{Ha09}.

We develop a statistical-mechanical model involving effective pair interactions between the cells that consist of hard-core repulsion and extra short-ranged soft-core repulsion beyond the hard core, whose length scale is roughly the same as the hard core. The model parameters are optimized to match the sample pair statistics in both direct and Fourier spaces, i.e., pair correlation function $g_2(r)$ and structure factor $S(k)$ (defined in the Materials and Methods Section). By doing this, the parameters are biologically constrained. For simplicity, we focus on modeling the patterns in late developmental stages. The generalization of the model to predict the more complex early-stage highly anisotropic cell patterning is deferred to a future work. We compute Voronoi statistics of the simulated configurations. In contrast with the aforementioned vertex-based models, our statistical-mechanical model does not explicitly involve interfacial energy functionals between cells that dictate the formation of cell shapes; nonetheless, our model predicts essentially the same polygonal shape distribution and size disparity of cells found in experiments as measured by Voronoi statistics. Moreover, our model is able to match other nontrivial unconstrained statistics such as the nearest-neighbor statistics and local number variance that measures density fluctuations of cell centroids across length scales. This is a testament to the power and novelty of the model. In addition, we find that the point patterns are not hyperuniform, which can be attributed to the fact that the skin must be pliable to deformations. This is to be contrasted with disordered hyperuniform photoreceptor-cell mosaics in avian retinae \cite{Ji14}, which must be mechanically rigid to sample light uniformly. The latter situation is consistent with the fact that disordered hyperuniformity is a property of a wide class of maximally random jammed states \cite{Za09, Do05b, Ji11b, Za11b, Za11d, At13, Ch14a, Dr15} that are indeed infinitely rigid. Disordered hyperuniform systems suppress large-scale density fluctuations like perfect crystals, and yet are statistically isotropic with no Brag peaks like liquids or glasses. They can be considered to be exotic states of matter that lie between the crystal and liquid states \cite{To03, Ji14}.

Since the correlations of cell centroids reflect the spatial distributions of cell nuclei, cytoplasm and membrane phases, and properties that ultimately affect the biological functions of the skin, our results should be able to provide new benchmarks that could be used to distinguish between normal and pathological skin tissues \cite{Ra15}. Moreover, our statistical-mechanical model enables us to generate tissue microstructure at will for further analysis, including the study of other structural and topological features (including anisotropic cell-packing patterns at the early stage, planar cell polarity, and ordered pattern in nonproliferating tissue, to name a few) as well as physical properties (including transport, elastic and viscoelastic properties). In addition, there are many ways in which our statistical-mechanical model might be extended to enhance our understanding of complex biological phenomena and processes in skin \cite{In06, Su09, Sa11, Be11}, and applied to the design and generation of artificial (or synthetic) tissues, which we will discuss in detail later.

The rest of the paper is organized as follows: in Materials and Methods, we first introduce the procedures to obtain and process images of epidermal patterns in mouse skin in different stages of embryonic development. We then introduce the definitions of the various structural descriptors employed in this work. In Results and Discussion, we employ these structural descriptors to characterize the evolving epidermal patterns and construct a computational model of the point patterns derived from the cell centroids in the late-stage experimental sample. In Conclusions, we offer concluding remarks, and propose directions in which our work can be extended.

\section*{Materials and Methods}
\subsection*{Whole mount immunostaining and image acquisition}
All experiments were performed on E12.5 or E14.5 embryos obtained by mating wild type \textit{C57BL/6J} (Jackson Laboratories) mice. Mice were handled and housed according to the approved Institutional Animal Care and Use Committee (IACUC) protocols of the Princeton University. For immunostaining, E12.5 or E14.5 embryos were dissected in phosphate-buffered saline (PBS) and fixed for 1 hour at room temperature in $4\%$ paraformaldehyde. Dissected backskins were permeabilized and blocked with PBS, $0.2\%$ Triton X-100, $2.5\%$ normal goat serum, and $2.5\%$ normal donkey serum for 1 hour at room temperature. Primary antibodies were incubated for overnight at $4\,^{\circ}\mathrm{C}$. Samples were then washed for three times 30 minutes in PBT ($0.2\%$ Triton X-100, diluted with PBS) at room temperature. Secondary antibodies, Phalloidin, and Hoechst were incubated for 3 hours at room temperature. Samples were washed in PBS at room temperature and mounted in Fluoro-Gel mounting medium (Electron Microscopy Sciences, Cat. $\#$ 17985-30). The following antibodies were used for immunofluorescence: rat anti-E-cadherin (DECMA-1, 1:500, Thermo Pierce, Cat. $\#$ MA1-25160), Rhodamine Phalloidin (1:1000, Cytoskeleton, Inc.) and secondary antibodies conjugated to AF-488 or AF-647 (1:1000, Invitrogen). Immunostained samples were imaged using an inverted Nikon A1R-Si confocal microscope, on a Nikon Eclipse Ti stand (Nikon Instruments) equipped with a GaASP detector. Images were acquired with a $40\times$ oil (N. A. 1.3) objective.

\subsection*{Image processing and extraction of cell centroids}
\begin{figure}[H]
\begin{center}
$\begin{array}{c}\\
\includegraphics[width=0.80\textwidth]{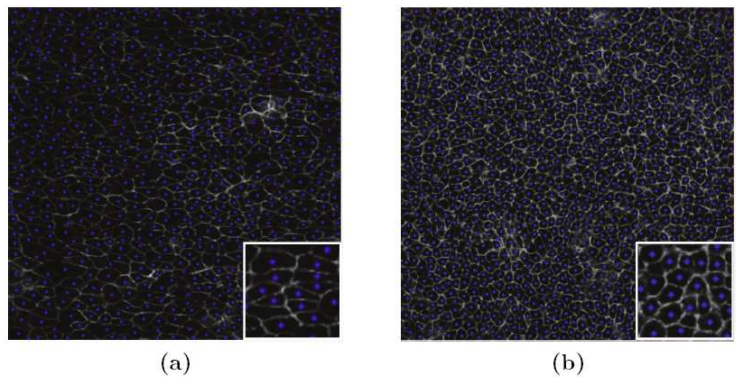}
\end{array}$
\end{center}
\caption{Representative disordered epidermal patterns in the basal layer of mouse skin at early (a) and late (b) stages of embryonic development after processing. The anterior-posterior axis runs vertically, while the medial-lateral axis runs horizontally. There are about 668 and 1085 cells (each of which has a linear size of about 7 $\mu m$) in the early and late stages, respectively. The cell membranes, and extracted cell centroids are indicated in light gray, and blue, respectively. The insets show blowup of a portion of the original images, which are edited using the Adobe Photoshop software.} \label{fig_1}
\end{figure}

The Packing Analyzer Software Package \cite{Ai10} is employed to identify all the cell membranes and then our in-house code \cite{Za11a} to ``dilate'' the membrane of each cell, i.e., convert all the intracellular pixels bordering the cell membrane into ``membrane'' pixels. This process separates the intracellular region of each cell from one another, allowing us to determine the position of each individual cell centroid (shown as blue dots in Fig. \ref{fig_1}) by averaging over the positions of the pixels of the associated intracellular region. The pixels corresponding to the intracellular region of a cell form a ``cluster'', which is connected by a path entirely in the intracellular region of that cell and identified using the ``burning'' algorithm \cite{St94, Ji09}. Note that the epidermis is slightly curved and the cells are not always in exactly the same plane, so pairs of epidermal cells, when they are projected onto the same plane in the imaging process, sometimes appear to be closer to one another than they are in actuality. The cell membranes are sometimes blurred due to noises present in the labeling and imaging processes, leading to errors in the subsequent step of identifying cell membranes. These effects introduce small errors in the final extracted positions of the cell centroids but do not affect the overall statistics, especially on large length scales. In addition, while the experimentally obtained tissue samples are relatively large, periodic boundary conditions are applied to the point patterns associated with the cell centroids to minimize boundary effects and approximate the infinite system.

\subsection*{Structural descriptors of epidermal patterns}
For point configurations derived from the epidermal cell centroids under periodic boundary conditions, we consider a variety of different types of lower-order correlation functions that are sensitive in picking up structural information across length scales \cite{To02, To03, Za09, Za11b, Za11c, Dr15}. A basic quantity of a point configuration is the pair correlation function $g_2({\bf r})$, which is proportional to the probability density function of finding two centers separated by the vector displacement ${\bf r}$ \cite{To02}. In practice, $g_2({\bf r})$ is computed via the relation
\begin{equation}
\label{eq_1} g_2({\bf r}) = \frac{\langle N({\bf r})\rangle}{\rho r\Delta r\Delta\theta},
\end{equation}
where $\langle N({\bf r})\rangle$ is the average number of particle centers that fall into the area element at a vector displacement ${\bf r}$ from a central particle center (arbitrarily selected and averaged over all particle centers in the system), $r\Delta r\Delta\theta$ is the finite differential area element ($r$ is the distance from the origin and $\theta$ is the polar angle), and $\rho$ is the number density of the point patterns \cite{To02, At13}. The structure factor $S({\bf k})$ is the Fourier counterpart, defined by
\begin{equation}
\label{eq_2} S({\bf k})= 1+ \rho {\tilde h}({\bf k}),
\end{equation}
where ${\tilde h}({\bf k})$ is the Fourier transform of the total correlation function $h(\mathbf{r}) = g_2(\mathbf{r})-1$ \cite{To02}, and ${\bf k}$ is the wavevector. The physical reason why one subtracts unity from $g_2$ to get the total correlation $h(r)$ is because in a disordered system without any long-range order, $h(r)$ will tend to zero for large $r$. Therefore, when $h(r)$ deviates from zero, it is a measure of correlations, both positive and negative. Note that Eq. \ref{eq_2} implies that the forward scattering contribution to the diffraction pattern is omitted. For computational purposes, the structure factor $S({\bf k})$ can be obtained directly from the particle positions ${\bf r}_j$, i.e.,
\begin{equation}
\label{eq_3}
S({\bf k}) = \frac{1}{N} \left |{\sum_{j=1}^N \exp(i {\bf k} \cdot
{\bf r}_j)}\right |^2 \quad ({\bf k} \neq {\bf 0}),
\end{equation}
where $N$ is the total number of points in the system \cite{Za09, At13, Ch14a}. The trivial forward scattering contribution (${\bf k} = 0$) in Eq. \ref{eq_3} is omitted, which makes Eq. \ref{eq_3} completely consistent with Eq. \ref{eq_2} in the ergodic infinite-system limit. For infinite point configurations in $d$-dimensional Euclidean space, they are hyperuniform if
\begin{equation}
\label{eq_4} \lim_{k\rightarrow0}S(k)=0,
\end{equation}
which implies that (normalized) infinite-wavelength density fluctuations of the system vanish \cite{To03, Za11b, Za11c, Ji11b}. In general, the spectral density ${\tilde \chi}({\bf k})$ instead of $S({\bf k})$ should be employed to characterize systems with particle size disparity \cite{Za11b, Za11d}; however, in the present case the size contrast between different cells is relatively small and we avoid making assumptions about cell sizes and shapes by using $S({\bf k})$.

To characterize local topology and size disparity of epidermal cells, we construct Voronoi tessellations of the point patterns associated with the cell centroids and compute statistics of the Voronoi cells \cite{To02}. A Voronoi cell is the region of space closest to a point than to any other point in the patterns \cite{To02}. A Voronoi tessellation is a tessellation of the space by the Voronoi cells \cite{To02}. The number of neighbors $n$ of a Voronoi cell is a discrete topological property of an epidermal cell, and the area of a Voronoi cell $A$ quantifies the size of an epidermal cell. The normalized standard deviation $c_n\equiv\sigma_n/\langle n\rangle$ of $n$ quantifies the local topology disparity of epidermal cells, where $\sigma_n$ and $\langle n\rangle$ are the standard deviation and ensemble average of $n$ \cite{Qu08, Mi12, Ki15}. The normalized standard deviation $c_A\equiv\sigma_A/\langle A\rangle$ of $A$ quantifies the size disparity of epidermal cells, where $\sigma_A$ and $\langle A\rangle$ are the standard deviation and ensemble average of $A$ \cite{Qu08, Mi12, Ki15}.

We also utilize the following three structural descriptors to characterize the density fluctuations of cell centroids and spatial correlations between nearest neighbors:
\begin{itemize}
\item $\sigma^2(R)$ -- Local number variance $\equiv$ variance of the number of particle centers that fall into the observation window with radius $R$ randomly placed in the system;
\item $H_P(r)$ -- Nearest-neighbor probability density function $\equiv$ probability density of finding the center of the nearest particle at a distance between $r$ and $r + dr$ from a given particle center; and
\item $E_P(r)$ -- Exclusion probability $\equiv$ probability of finding a circular cavity of radius $r$ centered at some arbitrary particle center empty of other particle centers.
\end{itemize}
To compute $\sigma^2(R)$, we randomly place circular observation windows with radius $R$ in the system under the constraint that the windows should fall entirely within the simulation box to avoid boundary artifacts \cite{Ji14, To03, Za11c}. Also, the largest radius of the window that one can sample must be smaller than half of the box length, otherwise density fluctuations are artificially diminished \cite{Dr15}. We count the number of cell centroids $N(R)$ that fall into the observation window, which is a random variable. The variance associated with $N(R)$ is denoted by $\sigma^2(R)\equiv\langle N(R)^2\rangle-\langle N(R)\rangle^2$, which measure local density fluctuations of cell centroids within a window of radius $R$. We note that $\sigma^2(R)$ is another quantity that can be employed to check hyperuniformity. For two-dimensional disordered hyperuniform systems, $\sigma^2(R)$ grows more slowly than the area of the circular observation window, i.e., more slowly than $R^2$ \cite{To03, Za09}. To compute $H_P(r)$, we bin the distances between each cell and its nearest neighbor and divide the number in each bin by the total number of cells and the bin size \cite{To02, Ji11}. From $H_P(r)$, we can compute the associated complementary cumulative distribution function $E_p(r)$ \cite{To02, Ji11} via the
\begin{equation}
\label{eq_5} E_P(r) = 1 - \int_{0}^{r}H_P(x)\mathrm{d}x.
\end{equation}

\section*{Results and Discussion}
\subsection*{Structural characterization of epidermal patterns}
\begin{figure}[H]
\begin{center}
$\begin{array}{c}\\
\includegraphics[width=0.80\textwidth]{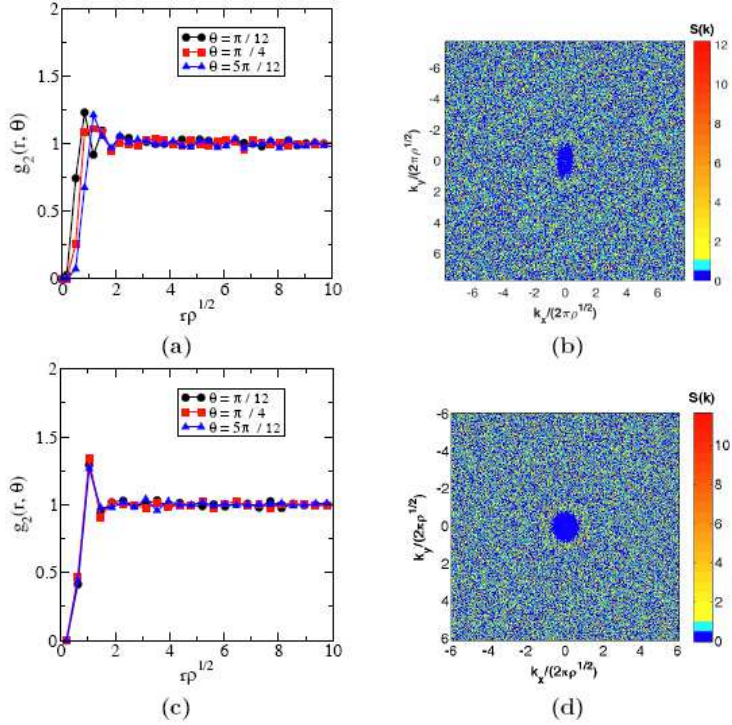}
\end{array}$
\end{center}
\caption{(a-b) Directional pair correlation function $g_2({\bf r})$ and structure factor $S({\bf k})$ of epidermal cell centroids in the early stages. Note that the first peak of $g_2({\bf r})$ shifts towards large $r$ and $g_2({\bf r})$ decays to its long-range asymptotic value of unity more slowly as $\theta$ increases, where $\theta$ is the angle between the sampling and vertical directions. The blue region in the vicinity of the origin, where $S({\bf k})$ is small, possesses an ``elliptical'' shape, implying larger density fluctuations on large scales in the horizontal direction. This anisotropy is due to the apparent ``stretching'' of the cells and the associated larger exclusion volume in the horizontal direction. (c-d) Directional pair correlation function $g_2({\bf r})$ and structure factor $S({\bf k})$ of epidermal cell centroids in the late stages, which show no directional dependence.} \label{fig_2}
\end{figure}

In the early stages, cells are stretched in the medio-lateral direction (horizontal in Fig. \ref{fig_1}), which promotes uniaxial growth by biasing cell divisions and cell rearrangements along the medio-lateral axis \cite{Aw16}. These events ultimately relax tissue anisotropy, and at later stages cells become more isotropic (circular) in shape \cite{Aw16}, as seen from Fig. \ref{fig_1}. To quantify the degree of statistical anisotropy and spatial correlations, we employ the aforementioned directional pair correlation function $g_2({\bf r})$ and the structure factor $S({\bf k})$, as shown in Fig. \ref{fig_2}. We find the pair statistics of the early-stage patterns show a strong directional dependence, i.e., statistical anisotropy. Note that the first peak of $g_2({\bf r})$ shifts towards large $r$ and $g_2({\bf r})$ decays to its long-range asymptotic value of unity more slowly as $\theta$ increases, where $\theta$ is the angle between the sampling and \textit{vertical} directions. The blue region in the vicinity of the origin, where $S({\bf k})$ is small, possesses an ``elliptical'' shape, implying larger density fluctuations on large scales in the horizontal direction. This anisotropy is due to the fact that cells at this stage are stretched along the medial-lateral axis and thus possess larger exclusion volumes \cite{Kr10, Ji14} in the horizontal direction. By contrast, both $g_2({\bf r})$ and $S({\bf k})$ in the late stages of development show no directional dependence, indicating that the corresponding point patterns of epidermal cells are statistically isotropic. This change from anisotropy to isotropy in the latter stages reflects the tissue relaxation resulting from oriented cell divisions, and provides a quantitative characterization of the morphological events that accompany global cell polarization as the epidermis becomes increasingly planar polarized \cite{De11, Aw16} and shifts from a program of planar, uniaxial growth to stratification \cite{Le05, Aw16}.
\clearpage

\subsection*{Statistical-mechanical model of late-stage epidermal patterns}
Using the aforementioned structural information revealed by the pair statistics, we embark on developing a statistical-mechanical model that accounts for packing effects. A packing in $d$-dimensional Euclidean space $\mathbb{R}^d$ is traditionally defined as a large collection of nonoverlapping solid objects (particles). Here we generalize the concept to allow for overlap of cells in order to account for the deformability of cell membranes \cite{Be05}. Our first goal is to devise a statistical-mechanical model of cells interacting via a particular choice of parameterized effective pair potential functions with parameters optimized to generate patterns that best fit the experimental configuration of cell centroids at late developmental stages. We note that our late-stage statistical-mechanical model can be generalized to predict the more complex early-stage highly anisotropic cell patterning by tuning the interactions between cells, but such studies are deferred to a future work.

\begin{figure}[H]
\begin{center}
$\begin{array}{c}\\
\includegraphics[width=0.60\textwidth]{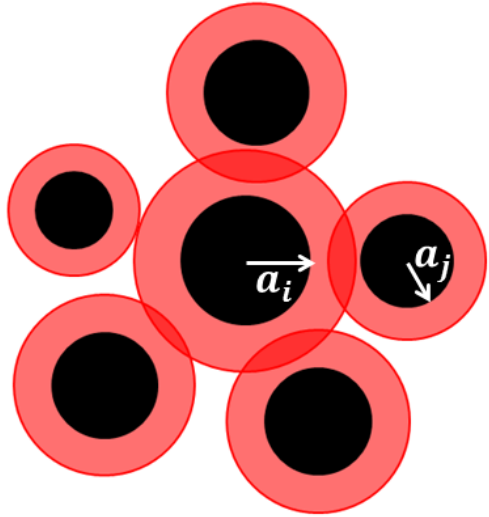}
\end{array}$
\end{center}
\caption{Illustration of the hard- and soft-core interactions in a system containing cells with different sizes. The solid black disks and larger concentric red (or light gray in the print version) circles illustrate the hard- and soft-core repulsions, respectively. The hard-core radius of cell $i$ is denoted by $a_i$.} \label{fig_3}
\end{figure}

We first measure the inradius distribution of the Voronoi cells associated with the epidermal cell centroids and use this as the starting point to estimate the size distribution of particles in our model. Then we use the specific functional forms of the measured pair statistics [$g_2(r)$ and $S(k)$] to suggest a class of effective cell-cell interactions that consist of two isotropic parts: a hard-core repulsion and a soft-core repulsion beyond the hard core, with a cutoff that is roughly of the same scale as the hard core (see Fig. \ref{fig_3}). From a qualitative observation of the cells, the nucleus is nearly always circular to oval, whereas the cytoplasm and membrane can be stretched and shaped into countless configurations. Therefore, it seems reasonable to assume that hard-core and soft-core interactions mimic nucleus and cytoplasm/membrane. Moreover, we expect that the larger is the area of the stiffer part of a cell, the larger is the range of the effective hard core of the cell. In addition, we expect that the ratio of the soft-core and hard-core radii is qualitatively proportional to the ratio of the area of the cell and that of the stiffer part of the cell. Importantly, the latter quantity is experimentally measurable. However, we stress that these are effective interactions and the range of the hard-core repulsion is not necessarily the same as the range of the stiffer part of the cell. Specifically, the strength of the hard-core repulsion is characterized by the radius $a_i$ of a hard-disk exclusion region associated with a cell type $i$. This interaction imposes a nonoverlap constraint such that the distance between the cells $i$ and $j$ can not be smaller than the sum of their hard-core radii, which mimics the physical cell packing constraint. The effective packing fraction $\phi$ of the cells (i.e., the fraction of space covered by the hard-disk exclusion regions of cells), is related to the size distribution of the cells via
\begin{equation}
\label{eq_6} \phi = \frac{1}{A_s}\sum_i N_i \pi(a_i)^2,
\end{equation}
where $A_s$ is the area of the system, and $N_i$ and $a_i$ are the number and radius of cells of type $i$, respectively. Note that high $\phi$ generally corresponds to low cell elasticity and motility. The range of the soft core is proportional to that of the hard core with the coefficient $\alpha > 1$ for every cell. Moreover, the pair potential $v(r_{ij})$ between cells $i$ and $j$ is given by
\begin{equation}
\label{eq_7} v(r_{ij}) = \left \{
\begin{array}{l@{\hspace{0.3cm}}c}
\infty, & r_{ij}\leq a_i + a_j
\\ \epsilon[-(\frac{r_{ij}-a_i-a_j}{\rho^{-1/2}})^\beta+[\frac{(\alpha-1)(a_i + a_j)}{\rho^{-1/2}}]^\beta], & a_i + a_j\leq r_{ij}\leq\alpha(a_i + a_j)
\\ 0, & r_{ij}\geq \alpha(a_i + a_j)
\end{array} \right .
\end{equation}
where $r_{ij}$ is the distance between cell $i$ and $j$, $a_i$ and $a_j$ are the hard-core radii of cell $i$ and $j$, $\rho$ is the number density of the simulated system, and parameters $\alpha>1$, $\beta>0$, $\epsilon$ set the scale of the interaction energy. The energy of the system $E$ is the sum of these effective pairwise repulsions, i.e.,
\begin{equation}
\label{eq_8} E=\sum_{i<j}v(r_{ij}).
\end{equation}

\begin{figure}[H]
\begin{center}
$\begin{array}{c}\\
\includegraphics[width=0.80\textwidth]{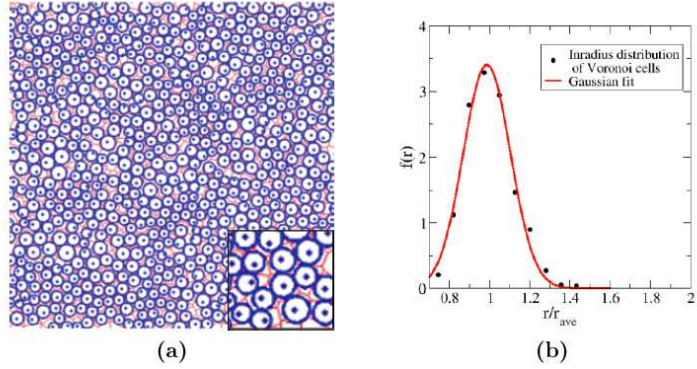}
\end{array}$
\end{center}
\caption{(a) Illustration of Voronoi cells of the epidermal patterns at late stage of embryonic development and their associated incircles. The inset shows a blowup of a portion of the original image, which is edited using the Adobe Photoshop software. (b) Inradius distribution of the Voronoi cells shown in (a). Cells within a distance of the sample boundary $\delta\leq0.1L$ (where $L$ is the linear size of the sample) are excluded to avoid boundary artifacts when calculating the distribution, and these cells are not shown in (a). Note that the inradius distribution can be well approximated by a normal distribution with a minimum cutoff.} \label{fig_4}
\end{figure}

We consider the inradius of the Voronoi cell associated with an epidermal cell centroid [see Fig. \ref{fig_4}(a)] as a rough measure of the actual size of the epidermal cell. We exclude cells within a distance of the sample boundary $\delta\leq0.1L$ (where $L$ is the linear size of the sample) to avoid boundary artifacts and compute the inradius distribution of Voronoi cells. We find that the distribution could be approximated well by a normal distribution with a minimum cutoff, as shown in Fig. \ref{fig_4}(b). Therefore, we employ a multi-component system with finite number of cell types and a cell-size distribution that approaches the normal distribution with a minimum cutoff as a starting point. We gradually increase the number of cell types in the system until we determine the minimal number of components that predicts the experimental patterns. We refer to this as a ``minimalist'' statistical mechanical model.

The parameters of the potential are optimized by minimizing the following objective function:
\begin{equation}
\label{eq_9} \Delta = \sum_r [g_2(r) - \overline{g}_2(r)]^2 + \sum_k [S(k) - \overline{S}(k)]^2.
\end{equation}
where $g_2(r)$ and $S(k)$ are the pair statistics of the simulated point patterns determined via Monte Carlo
methods from the aforementioned interactions at any point in the simulation, and $\overline{g}_2(r)$ and $\overline{S}(k)$ are the corresponding experimentally measured functions \cite{Ji14}. We note that, in principle, for infinite systems, the direct-space term and the Fourier-space term in Eq. \ref{eq_9} are equal by Parseval's theorem \cite{Fe56}. However, this is not true for finite-system simulations, where the summations over $r$ and $k$ are cut off at certain values. By including both terms in the objective function, we are able to capture both short and large-scale correlations in the experimental sample. The Monte Carlo algorithm (similar to the one in Ref. \cite{Ji14}) that we employ to generate different point patterns, which involves iterating ``growth'' and ``equilibration'' steps, works as follows:

(1) \textit{Initialization}. In the beginning of the simulation, cells are generated in a square simulation box using the random-sequential-addition (RSA) process \cite{To10a} with a prescribed size distribution. Specifically, the cells are randomly, irreversibly, and sequentially added into the system under the constraint that their hard cores do not overlap with the hard cores of the cells already in the system \cite{To10a}. The initial packing fraction of epidermal cells $\phi_I$ is about $50\%$ of the RSA saturation density \cite{To10a}.

(2) \textit{Growth.} At each stage $n$, the radius $a_i$ of each cell is increased by the same small fractional amount such that no pairs of hard cores of cells overlap. This leads to an increase of the packing fraction $\phi_n$ at this stage by an amount of about $1\%$-$3\%$. Then the cells are allowed to randomly move a direction and prescribed maximal distance such that no pairs of hard cores overlap for a certain number of movements ($\approx 1000$ per cell). Note that in this ``growth'' step, the extra soft repulsions between cells are turned off.

(3) \textit{Equilibration}. At the end of the ``growth'' step, the soft interactions are then turned on, and the system is allowed to equilibrate at fictitious ``temperature'' $T$ subject to nonoverlap conditions. Specifically, each cell is allowed to randomly move within a prescribed maximal distance from its old position and the trial move is accepted with the probability
\begin{equation}
\label{eq_10} p_{acc}(old\rightarrow new) = \textnormal{min}\{1, \textnormal{exp}(-\frac{E_{new}-E_{old}}{k_BT})\},
\end{equation}
where $E_{old}$ and $E_{new}$ are the energies of the system before and after the trial move as defined in Eq. \ref{eq_8}.

(4) \textit{Statistics}. After the equilibration process, structural statistics of the resulting configuration of cell centroids are obtained and compared to the corresponding experimental data.

(5) The growth and equilibration steps described in the bullet items (2) and (3), respectively, are repeated until $\phi_n$ reaches a prescribed value $\phi_F$. Specifically, the configuration obtained by equilibration at stage $n$ is used as the starting point for the growth step at stage $n+1$.

We optimize the parameters of the potential, fictitious ``temperature'' $T$ of the system, size distribution and final packing fraction $\phi_F$ of the cells in order to generate equilibrium ``liquid'' configurations that minimizes $\Delta$, i.e., best matches the functions $g_2(r)$ and $S(k)$. We find that a four-component system with $N_1/N_2/N_3/N_4=324:437:265:59$ and $a_1/a_2/a_3/a_4=0.78:1.0:1.22:1.44$ generates the best patterns, where $N_i$ and $a_i$ are the number and radius of cells of type $i$, respectively. We note that further increasing the number of components in the system beyond this four-component system does not significantly improve the results. We find the corresponding optimized parameter values to be $\alpha=1.5$, $\beta=2.0$, $k_B T/\epsilon=0.054$, and $\phi_F = 0.485$, where $k_B$ is the Boltzmann constant.

\begin{figure}[H]
\begin{center}
$\begin{array}{c}\\
\includegraphics[width=0.80\textwidth]{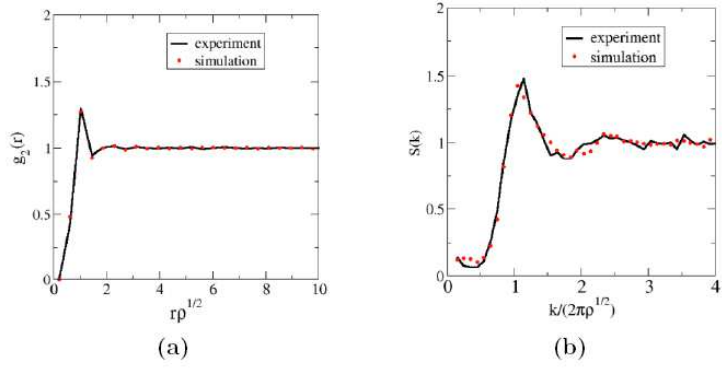}
\end{array}$
\end{center}
\caption{Experimentally measured and simulated (averaged over five configurations) pair correlation function $g_2(r)$ (a) and structure factor $S(k)$ (b) of epidermal patterns, which are in good agreement with one another.} \label{fig_5}
\end{figure}

Figure \ref{fig_5} shows the experimentally measured $g_2(r)$ and $S(k)$ and those of the final simulated point patterns (averaged over five configurations) with the minimal $\Delta_{min} < 0.10$. As one can see, our statistical-mechanical model with two-scale effective pair interactions between cells captures both the local and large-scale correlations in the actual system. It is also noteworthy that $g_2(r)$ approaches its large-$r$ asymptotic value of unity very quickly and $S(k)$ possesses a nonzero value at $k=0$. These results demonstrate that while large-scale density fluctuations are suppressed, the patterns are not hyperuniform, which distinguishes this epidermal system from hyperuniform avian retinal epithelial mosaics \cite{Ji14}. A reasonable explanation for this contrast between the two systems is that the former must be pliable to deformations, whereas the latter must be rigid to sample light uniformly. Note that hyperuniformity has been shown to arise when a variety of particle packings in two and three dimensions are driven to maximally random jammed (mechanically rigid) states \cite{Za09, Do05b, Ji11b, Za11b, Za11d, At13, Ch14a, Dr15}. Indeed, the elastic moduli are infinite. Moreover, from equilibrium statistical mechanics, it is well-known that when $S(0)$ of a system is not zero, i.e., the system is not hyperuniform, the system is less constrained and typically fluid-like \cite{To02}, which in the context of biological cells one can qualitatively think of as being malleable with relatively small elastic moduli.

\begin{figure}[H]
\begin{center}
$\begin{array}{c}\\
\includegraphics[width=0.80\textwidth]{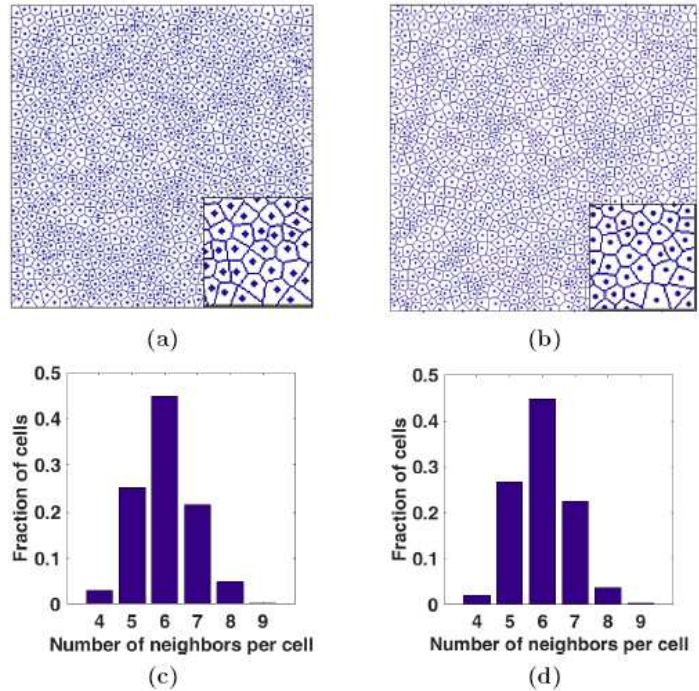}
\end{array}$
\end{center}
\caption{Experimentally obtained (a) and representative final simulated (b) point patterns of epidermal cell centroids and their associated Voronoi cells. The insets show blowup of a portion of the original images, which are edited using the Adobe Photoshop software. The polygonal shape distribution of the Voronoi cells (or the number of neighbors per cell) associated with the experimental and simulated (averaged over five configurations) point patterns are shown in (c) and (d), respectively. Note that the point patterns in (a) is a mapping from the epidermal patterns in Fig. \ref{fig_1}(b). Also, cells within a distance of the sample boundary $\delta\leq0.1L$ (where $L$ is the linear size of the sample) are excluded to avoid boundary artifacts when calculating the distributions in (c) and (d). These two point patterns and their associated Voronoi cells look indistinguishable from each other visually. It is noteworthy that our model predicts essentially the same distribution of cell shapes as that in actuality without explicitly considering interfacial energy between cells.} \label{fig_6}
\end{figure}

\begin{figure}[H]
\begin{center}
$\begin{array}{c}\\
\includegraphics[width=0.80\textwidth]{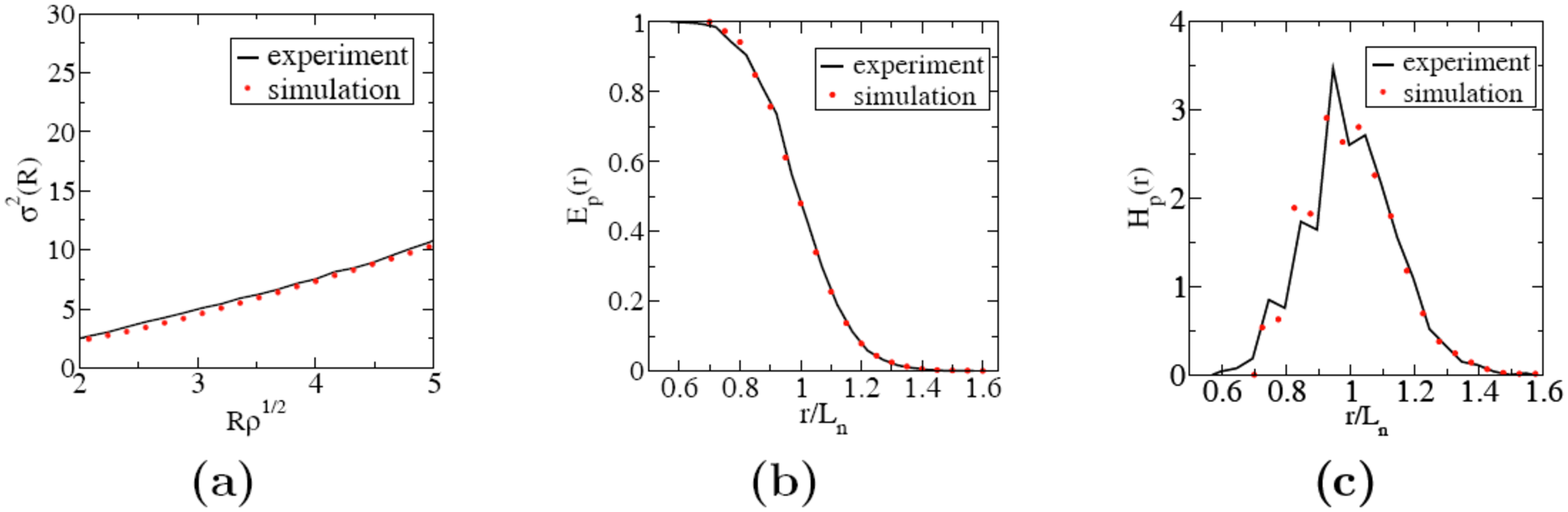}
\end{array}$
\end{center}
\caption{Experimentally measured and simulated (averaged over five configurations) local number variance $\sigma^2(R)$ (a), ``particle'' exclusion probability $E_P(r)$ (b), and ''particle'' nearest-neighbor probability density function $H_P(r)$ (c) for epidermal patterns, where $L_n$ is the mean nearest-neighbor distance. Our simulated configurations matches the experimental sample well as measured by all of these statistics.} \label{fig_7}
\end{figure}

The final simulated configurations are endowed with structural descriptors, beyond the targeted ones [$g_2(r)$ and $S(k)$] that are in excellent agreement with the ones associated with the experimentally derived image [Fig. \ref{fig_1}(b)]. Specifically, we construct Voronoi tessellations of experimental and simulated point patterns and compute the corresponding polygonal shape (i.e., number of neighbors) distributions of the resulting Voronoi cells, as shown in Fig. \ref{fig_6}. Note that in this calculation, we exclude cells within a distance of the sample boundary $\delta\leq0.1L$ (where $L$ is the linear size of the sample) to avoid boundary artifacts. Voronoi tessellations associated with experimental and simulated (averaged over five configurations) systems show similar distributions of polygonal cell shapes, with most of the cells possessing six neighbors. Moreover, we compute the aforementioned $c_n$ and $c_A$ of the Voronoi cells, and the results are shown in Table \ref{table_1}. We find a good agreement on the values of $c_n$ and $c_A$ between the experimental and simulated patterns. This is remarkable since our statistical-mechanical model does not explicitly consider interfacial energy between cells; nonetheless, the distribution of cell shapes is recovered from simple effective interactions between cell centroids. Furthermore, we compute $\sigma^2(R)$, $H_P(r)$, $E_P(r)$ of the simulated configurations (averaged over five configurations), which are in very good agreement with the corresponding quantities of the experimental configurations, as shown in Fig. \ref{fig_7}. This indicates that we again accurately capture both local and large-scale spatial correlations associated with experimental system. Note that $\sigma^2(R)$ grows as $R^2$, which again demonstrates that point patterns of cell centroids are not hyperuniform. In summary, the above results demonstrate that our statistical-mechanical model, as a complementary approach to vertex-based models \cite{Ho04, Fa07, Hi08, St10}, can be used to perform \textit{in-silico} predictions of epidermal patterns.

\begin{table}[H]
\caption{Topology and size disparity of late-stage experimental and simulated (averaged over five configurations) epidermal patterns. The quantities $c_n$ and $c_A$ are normalized standard deviations of the number of neighbors and area of Voronoi cells, respectively.}
\begin{center}
\begin{tabular}{{c}{c}{c}} \\ \hline\hline
 & Experimental patterns & Simulated patterns \\
\hline
$c_n$ & 0.1488 & 0.1444 \\
\hline
$c_A$ & 0.1810 & 0.1924 \\
\hline\hline
\end{tabular}
\end{center}
\label{table_1}
\end{table}

\section*{Conclusions}
In this paper, we first employed directional pair correlation function $g_2({\bf r})$ and structure factor $S({\bf k})$ to quantitatively characterize evolving epidermal patterns extracted from histological images of mouse skin. We find that the pair statistics of the patterns in the early stages of embryonic development show structural directional dependence, which is a reflection of the fact that cells are stretched along the medial-lateral axis to promote uniaxial growth \cite{Aw16}. By contrast, in the late stages the patterns possess statistical isotropy, which likely results from the oriented divisions and cell rearrangements that relax tissue strain \cite{Aw16}. Increased isotropy accompanies global cell polarization and a shift towards stratifying divisions, which may be the functional consequences of these morphological changes. By matching $g_2(r)$ and $S(k)$, we constructed a minimalist four-component statistical-mechanical model involving effective isotropic short-ranged repulsive pair interactions between cells to predict the late-stage patterns. The ascertained model parameters were biologically constrained. We constructed the Voronoi tessellations of the simulated configurations and computed the statistics of the Voronoi cells. In contrast with many vertex-based models, our statistical-mechanical model does not explicitly consider interfacial energy between cells; nonetheless, our model predicts essentially the same polygonal shape distribution (in particular its normalized standard deviation $c_n$), and normalized area standard deviation $c_A$ of Voronoi cells found in experiments. These results demonstrate that our statistical-mechanical model effectively captures key cell mechanics, including local structural features in epidermis.

As a testament to its power and novelty, our model is able to match other nontrivial unconstrained statistics such as the nearest-neighbor probability density function $H_P(r)$ and its associated complementary cumulative distribution function $E_P(r)$, and the local number variance $\sigma^2(R)$. In addition, we find that the point patterns of cell centroids are not hyperuniform, which we conclude is consistent with the deformability of skin. This is to be contrasted with hyperuniform photoreceptor-cell mosaics in avian retinae, which must be rigid to sample light uniformly. Our results demonstrate that our statistical-mechanical model, as a complementary approach to vertex-based models, can be used to predict epidermal patterns.

Our results do not imply that there are effective two-scale pair interactions between epidermal cells operating during evolution. Rather the significance of our results lies in an ability to map the modeling task to the solution of statistical-mechanical problem of interacting particles and predict not only the specific late-stage epidermal patterns but a family of epidermal patterns that can be tuned by varying the parameters of the effective pair interactions. Future work will include the introduction of directional effective interactions to predict the epidermal patterns in the early stages that are statistically anisotropic. Moreover, changing the set of parameters alters the tissue structure, which ultimately affects the effective properties of the tissue. At the same time, the tissue structure is intimately related to the properties of the individual biological constituents, such as cell elasticity, tension as well as adhesion, which is a surface interfacial property. Thus, these cell properties are indeed reflected by the effective parameters in our statistical-mechanical model. Our biologically-constrained model enables us to generate tissue microstructure at will for further analysis, including the study of other structural and topological features as well as physical properties. Furthermore, our model could be extended to allow for nonspherical-particle shapes and incorporating local directional interactions between neighboring cells. This extension would allow one to capture a wider range of structural features, including planar cell polarity, which was not considered in this work. In addition, the image-processing techniques and various structural descriptors that we employ in this work can be readily applied to investigate spatial correlations of cells across length scales in other biological tissues, such as cucumis and \textit{Drosophila} epithelia \cite{Ki15}.

It is noteworthy that factors such as the cell growth rate, division frequency (cell-cycle length), asymmetry in cell division, or mechanical deformations are known to have the ability to alter the size and shape of cells \cite{Gi15}. For example, dividing cells in \textit{Drosophila} epithelia tend to possess more neighbors and larger cell areas than nondividing cells \cite{He16}. Kidney epithelial cells alter size upon changes of rates of fluid flow in the nephron ducts and mechanical shear on the primary cilium \cite{Bo10}. By varying the size distribution of cells in our statistical-mechanical model, one can predict how epidermal cell packing patterns change upon varying the aforementioned factors.

Interestingly, epidermis may be thought of as multi-phase media consisting of the cell nuclei, cytoplasm and membrane phases, which possess effective bulk properties that are intimately related to biological function. This includes effective transport properties (e.g., diffusion coefficients and reaction rates), effective elastic moduli and viscoelastic characteristics, which depend on the properties of the individual phases as well as the structural correlation functions \cite{To02, Kr77, To91, To98, Gi98}. For example, the Poisson's ratio of all the phases are 0.5, and the Young's moduli of the nuclei, cytoplasm and membrane phases in human foreskin epithelial cells are 14 kPa, 37 kPa and 0.57 kPa, respectively, as measured by in-vitro atomic force microscopy experiments \cite{Be05}. The characteristic time for viscous dissipation in the cell membrane is about 0.1 s and the viscosity of intracellular cytoskeleton is about 100 Pa$\cdot$s, respectively, reflecting the viscoelastic nature of epidermis \cite{Ka03}. Therefore, using some of the structural descriptors computed in this work, rigorous analytical and simulation approaches from heterogeneous materials \cite{To02}, including strong-contrast expansions \cite{To02}, upper and lower bounds on mechanical properties \cite{To02, Gi98}, and finite-element techniques \cite{To04, Ha12}, may be fruitfully applied to enhance our understanding of the effective physical properties of the tissue. Moreover, one can bring to bear cross-property relations \cite{To02}, which, to our knowledge, have not been applied in biological contexts. Cross-property relations are rigorously exact expressions that relate one effective property (e.g., diffusion coefficient or reaction rate) to a given measurement of a different effective property (e.g., fluid permeability or elastic moduli); see Refs. \cite{To90, Gi93, Gi96} and references therein. Such a program would enable one to understand how the effective properties and associated biological functions of the tissue change during morphogenesis, wound-healing and disease progression processes in skin \cite{In06, Su09, Sa11, Be11}.

Furthermore, our statistical-mechanical model can be fruitfully applied to design and generate artificial (or synthetic) tissues. Specifically, by purposely tuning the effective parameters, we can manipulate the microstructures of the tissues and construct novel artificial (or synthetic) skin tissues \cite{Br11} with desirable mechanical and transport properties, which can then be realized using for example 3D printing technologies \cite{Mu14}.

\section*{Author Contributions}
S.T. and D.D. designed the research, S.T., D.D., D.C. and W.Y.A. performed the research; S.T., D.D., D.C. and W.Y.A. contributed analytic tools, S.T., D.D. and D.C. analyzed the data, and S.T., D.D. and D.C. wrote the paper.

\section*{Acknowledgements}
We are grateful to Steven Atkinson and Dr. Sascha Hilgenfeldt for very helpful discussions. D.D. lab is supported by National Institutes of Health/National Institute of Arthritis and Musculoskeletal Skin Diseases grant R01AR066070. W.Y.A. holds an American Heart Association graduate student fellowship.

%\bibliography{paper}
% Figure legends
%\clearpage
%\section*{Figure Legends}
%\subsubsection*{Figure~\ref{fig:result_fig}.}
%Figure legend here.

% Figures, one per page (fig_1.eps and fig_1.pdf files must be present
% in the document directory)
%\clearpage
%\begin{figure}
 %  \begin{center}
      %\includegraphics*[width=3.25in]{fig_1}
 %     \caption{}
  %    \label{fig:result_fig}
  % \end{center}
%\end{figure}

% closing statement, nothing below matters
\end{document}